\documentclass[10pt,twocolumn,english,superscriptaddress,showpacs,floatfix,nofootinbib]{revtex4}
\usepackage[T1]{fontenc}
\usepackage[latin9]{inputenc}
\usepackage{color}
\usepackage{amsmath}
\usepackage{amssymb}
\usepackage{graphicx}
\usepackage{esint}

\makeatletter
\@ifundefined{textcolor}{}
{%
 \definecolor{BLACK}{gray}{0}
 \definecolor{WHITE}{gray}{1}
 \definecolor{RED}{rgb}{1,0,0}
 \definecolor{GREEN}{rgb}{0,1,0}
 \definecolor{BLUE}{rgb}{0,0,1}
 \definecolor{CYAN}{cmyk}{1,0,0,0}
 \definecolor{MAGENTA}{cmyk}{0,1,0,0}
 \definecolor{YELLOW}{cmyk}{0,0,1,0}
}

\usepackage{braket}\usepackage{babel}
\usepackage{gensymb}

\usepackage{dcolumn}
\usepackage{bm}
\usepackage{epstopdf}\usepackage{amsfonts}\usepackage{array}\usepackage{float}\usepackage{xspace}

\newcommand{\hh}{\mbox{H$_{2}^{+}$}}
\newcommand{\dd}{\mbox{D$_{2}^{+}$}}
\newcommand{\ttt}{\mbox{T$_{2}^{+}$}}

\usepackage{babel}

\usepackage{babel}

\makeatother

\usepackage{babel}
\begin{document}


\title{Even harmonic generation in isotropic media of dissociating homonuclear
molecules}

\author{R. E. F. Silva}

\email{rui.silva@uam.es}

\affiliation{\emph{Departamento de Qu\'{\i}mica, Universidad Aut\'onoma de Madrid, 28049
Madrid, Spain}}

\author{P. Rivi\`ere}

\affiliation{\emph{Departamento de Qu\'{\i}mica, Universidad Aut\'onoma de Madrid, 28049
Madrid, Spain}}

\author{F. Morales}

\affiliation{\emph{Max-Born-Institut, Max Born Strasse 2A, D-12489 Berlin, Germany}}

\author{O. Smirnova}

\affiliation{\emph{Max-Born-Institut, Max Born Strasse 2A, D-12489 Berlin, Germany}}

\author{M. Ivanov}

\affiliation{\emph{Max-Born-Institut, Max Born Strasse 2A, D-12489 Berlin, Germany}}

\affiliation{\emph{Department of Physics, Voronezh State University, Universitetskaya
pl., 1, Voronezh, Russia, 394036}}

\author{F. Mart\'{\i}n}

\email{fernando.martin@uam.es}

\affiliation{\emph{Departamento de Qu\'{\i}mica, Universidad Aut\'onoma de Madrid, 28049
Madrid, Spain}}

\affiliation{\emph{Instituto Madrile\~no de Estudios Avanzados en Nanociencia, 28049
Madrid, Spain}}

\affiliation{\emph{Condensed Matter Physics Center (IFIMAC), Universidad Aut\'onoma
de Madrid, 28049 Madrid, Spain }}

\pacs{42.65.Ky, 33.80.Rv, 34.50.Gb}
\begin{abstract}
{Isotropic gases irradiated by long pulses of intense IR light can
generate very high harmonics of the incident field. It is generally
accepted that, due to the symmetry of the generating medium, be it
an atomic or an isotropic molecular gas, only odd harmonics of the
driving field can be produced. {Here we show how the interplay of electronic
and nuclear dynamics can lead to a marked breakdown of this standard
picture: a substantial
part of the harmonic spectrum can consist of \textit{even} rather
than odd harmonics.  
We demonstrate the effect using ab-initio solutions of the time-dependent
Schr\"odinger equation for H$_{2}^{+}$ and its isotopes in full dimensionality. By means of a simple analytical model,
we identify its physical origin, which is the appearance
of a permanent dipole moment in dissociating homonuclear molecules,
caused by light-induced localization of the electric charge during
dissociation. The effect arises for sufficiently long laser pulses and the region of the spectrum where even harmonics
are produced is controlled by pulse duration.} Our results (i) show how the interplay of femtosecond
nuclear and attosecond electronic dynamics, which affects the charge
flow inside the dissociating molecule, is reflected in the nonlinear
response, and (ii) force one to augment standard selection rules found
in nonlinear optics textbooks by considering light-induced modifications
of the medium during the generation process.} 
\end{abstract}
\maketitle

\section*{Introduction\label{sec:Introduction-1}}

{Attosecond technology originated in nonlinear optics, with high
harmonic generation (HHG) being the fundamental physical process underlying
the generation of attosecond pulses \cite{RevModPhys.81.163,scrinzi2006attosecond}.
In the two decades since its inception, attosecond science has grown
rapidly \cite{murmane_atto_1,sansone2006isolated,murmane_atto_2}
with} applications in physics, chemistry \cite{lepine2014attosecond,Nature_jonh,smirnova2009high},
materials science \cite{cavalieri2007attosecond,schiffrin2013optical,ghimire2011observation}
and even biology \cite{calegari2014ultrafast,kraus2015measurement}.
High harmonic emission results from nonlinear response of a medium
to an intense laser field. Its basic mechanism was first described
in \cite{HHG_CORKUM,KULANDER_HHG,HHG_QUANTUM} (see also \cite{kuchiev1987atomic}).
After the intense laser field frees an electron from the ionic core,
the electron gains energy from the field and revisits the parent ion.
Radiative recombination converts the gained energy into high-frequency
radiation.

In isotropic atomic gases irradiated by long IR pulses, {the electron
round-trips between ionization and recombination are launched during
successive laser half-cycles. Mirror symmetry of the driving electric
field implies that these round trips are mirror images of each other.
Electrons revisiting the parent ion from opposite directions yield
emission bursts} with the same amplitude but opposite signs. As a
consequence, even-order harmonics interfere destructively and vanish,
while the odd-order harmonics interfere constructively \cite{dahlstrom2011quantum},
{leading to the spectral peaks at odd multiples of the fundamental
frequency, $\Omega_{n}=(2n+1)\hbar\omega_{{\rm IR}}$.}



A similar behavior is commonly expected for any isotropic medium,
such as an isotropic distribution of homonuclear diatomic molecules.
However, the physical picture underlying high harmonic generation
suggests that the expectation of odd-only harmonics requires that
the inversion symmetry holds during the whole interaction. This is
indeed the case if the molecular nuclei do not, or barely, move. However,
as predicted in recent theoretical work on H$_{2}^{+}$ \cite{lein2005attosecond,morales2014high,riviere2014time,bian2014probing},
mostly using low dimensionality models \cite{lein2005attosecond,morales2014high,riviere2014time},
new features may arise in the high harmonic spectrum if the nuclei
move significantly during {one or several laser half
cycles.} These include the reduction of the maximum (cutoff) energy
in the harmonic spectrum \cite{lein2005attosecond}, a modest red
shift of the harmonic peak positions \cite{bian2014probing}, suppression
of specific odd harmonics \cite{riviere2014time} and, as found more
recently, the appearance of weak even harmonics between the strong
odd harmonic peaks \cite{morales2014high}. Common wisdom suggests
that these features should disappear for long driving laser pulses,
restoring the expected symmetry of the overall process and balancing
contributions from the adjacent laser half-cycles.

\begin{figure*}[t]
\begin{centering}
\includegraphics[width=2\columnwidth]{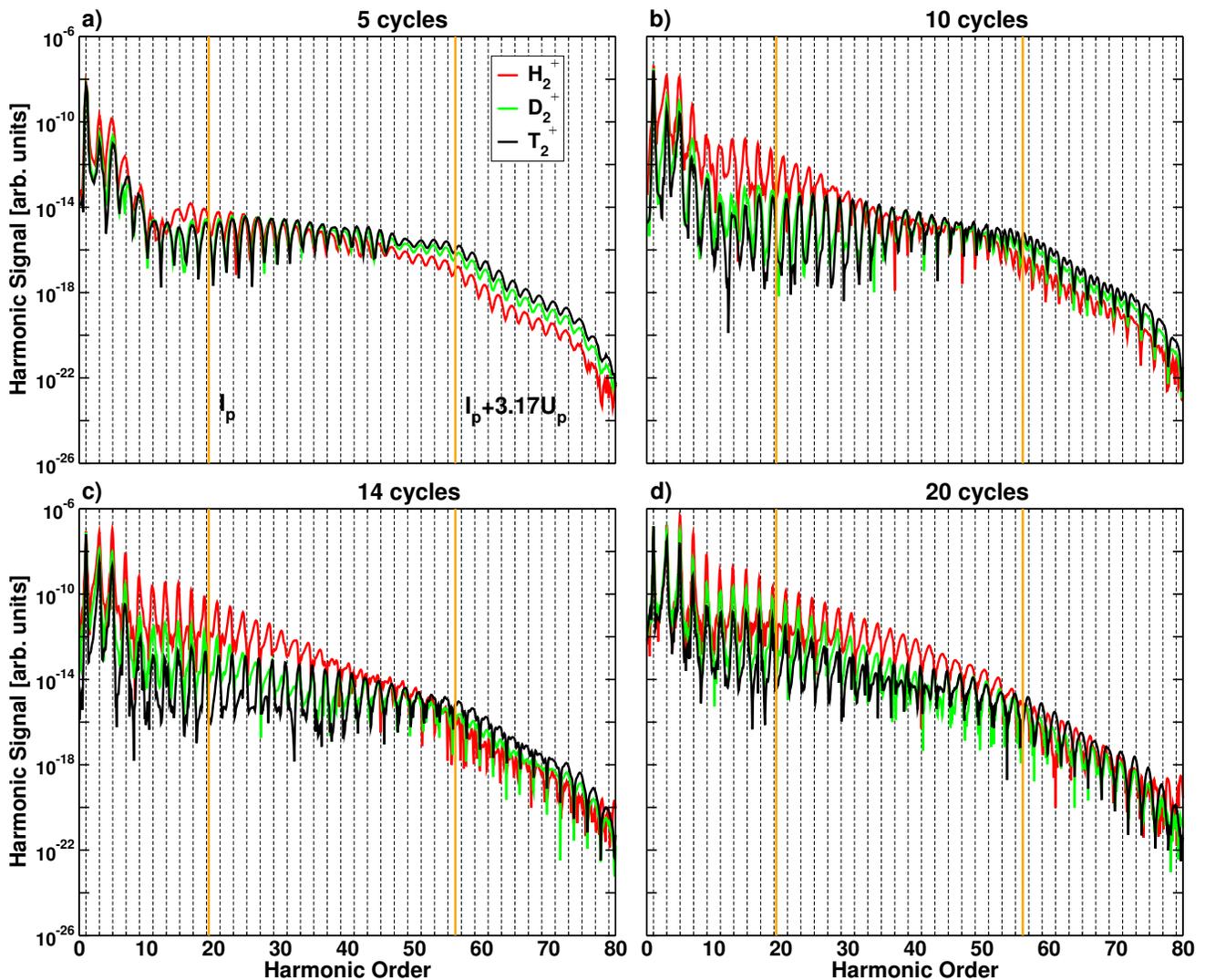} 
\par\end{centering}

\caption{\label{fig:HHG3D_5_cycles}HHG spectrum for a pulse of $800$ nm,
$I=3\times10^{14}$W/cm$^{2}$ and 5 (a), 10 (b), 14 (c) and 20 (d)
optical cycles for $\hh$, $\dd$ and $\ttt$. The dashed vertical
lines indicates odd harmonics. The thick vertical lines indicate the
ionization threshold (left line) and the cut-off energy obtained from
$E_{{\rm cutoff}}=I_{p}+3.17U_{p}$ (right line).}
\end{figure*}

{However, results of our study of high harmonic generation from H$_{2}^{+}$,
D$_{2}^{+}$ and T$_{2}^{+}$, using a full dimensionality model for
the electronic and vibrational degrees of freedom, contrast with this
expectation even for rather long laser pulses. We find that the nuclear
motion has an even more dramatic effect than anticipated in the previous
work. For sufficiently long pulses, the HHG spectrum of the lighter
molecules can exclusively consist of \textit{even} harmonics in the
plateau region. Furthermore, by changing the pulse duration, one can
control the region of the plateau where the even harmonics appear.
We unambiguously link the appearance of even harmonics to electron
localization on one side of the molecule. This induces a permanent
dipole during dissociation, even for pulses containing tens of laser
cycles. 
The linear Stark effect associated with the permanent dipole introduces
a relative phase between the two consecutive electron round trips
initiated in the adjacent laser half-cycles. As this phase approaches
$\pi$, the odd harmonics are replaced by the even ones. } 

%

\section*{HHG spectra\label{sec:HHG-spectra}}

Fig. \ref{fig:HHG3D_5_cycles} shows the calculated HHG spectra for
the three different isotopes of the $\hh$ molecule and different
pulse durations. For the shortest pulse, 5 optical cycles, our results
are {very close to our earlier results for low dimensionality models
\cite{morales2014high,riviere2014time}.} {The spectra from different
isotopes are generally very similar, with several broad even harmonic
lines between orders 36-40. The effect is independent of the isotope
and hence of the nuclear motion; it is associated with the frequency
chirp induced by the changing laser intensity between successive half-cycles
in an ultra-short pulse. As expected, the effect disappears for the
longer, 10-cycle pulse, and for the heaviest isotope, where the nuclear
motion is negligible. However, for the lightest isotope even harmonics
remain very prominent for 10, 14, and 20-cycle pulses. As one moves
to higher orders, the harmonic peaks in $\hh$ experience a red shift,
up to a point that only even harmonics are observed for high enough
orders. The spectral region where even harmonics dominate shifts to
higher orders with increasing pulse duration. For the 20-cycle pulse
even harmonics also appear for the D$_{2}^{+}$ isotope. Additionally,
for the longer pulses, the harmonics from D$_{2}^{+}$ and T$_{2}^{+}$
are strongly suppressed in the plateau region compared to H$_{2}^{+}$,
with rather dramatic modifications of the shape of the harmonic lines.
What could be the origin of even harmonics in H$_{2}^{+}$, their
shift with the pulse duration, and the emergence of two spectral regions
where they are seen for 20-cycle pulse (the plateau and the cutoff)?
What could be the origin of the combination of harmonic suppression
and the dramatic harmonic line shape modifications for T$_{2}^{+}$?}


{Two effects may be responsible. The first is electron localization,
prominent in molecular dissociation \cite{kling2006control,Nature_jonh}.
Electron localization breaks the symmetry of the system and hence
can lead to even harmonics, \cite{morales2014high,riviere2014time,smirnova2007anatomy}.
The second is the asymmetry introduced by molecular dissociation between
the raising and descending parts of the IR pulse \cite{bian2014probing},
in particular due to the shift in the characteristic ionization potential
$I_{p}$. As pointed out in Ref. \cite{bian2014probing}, this asymmetry
leads to the red shift. The analysis of the harmonic spectra for different
pulse durations and isotopes allows us to distinguish the contributions
of these two effects. We argue that both are important but have quite
different impact in multi-cycle pulses. While electron localization
controls the interference of the emission bursts from successive laser
half-cycles, the front-back asymmetry pertains to the longer time-scale
and hence leads to finer-scale modifications in the harmonic spectrum,
for long pulses.}

{We first analyze the impact of dissociation-induced asymmetry between
the raising and the falling edges of the pulse. As discussed above,
the harmonic spectrum is formed by the interference of emission bursts
produced during successive half-cycles. If the system response during
several successive half-cycles is nearly identical, odd harmonics
will form during the raising part of the pulse. The same would apply
at the falling edge of the pulse, even if the system has changed due
to dissociation. The interference of odd harmonics generated at the
front and at the back of the pulse will then lead to modification
of harmonic lineshapes, within their width. However, it will not turn
odd harmonics into even. If the contributions from the raising and
the falling parts of the pulse are phase-shifted by $\pi$ in some
spectral window, strong reshaping and suppression in this spectral
window may result. This is precisely what we find for the heaviest
isotope, T$_{2}^{+}$, for a 20-cycle pulse, between harmonics $33-45$:
overall suppression, strong modification of line-shapes, but prominent
odd harmonics nevertheless. To shift the harmonic lines on the scale
of $\hbar\omega_{{\rm IR}}$, where $\omega_{{\rm IR}}$ is the laser frequency, one has to modify the interference between
successive half-cycles. We now show that the generation of a permanent
dipole moment due to electron localization is the origin of such modification
and the appearance of even harmonics.}

\section*{Electron localization\label{sec:Electron-localization}}

\begin{figure}[t]
\begin{centering}
\includegraphics[width=0.9\columnwidth]{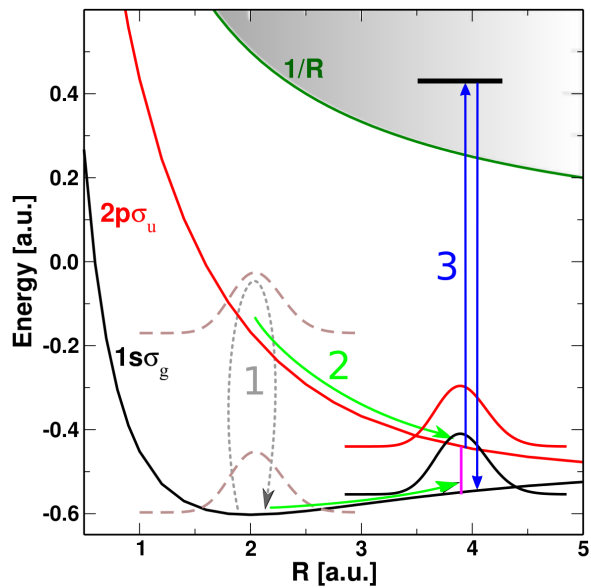} 
\par\end{centering}

\caption{\label{fig:HHG3D_Process}Sketch of the dynamics leading to electron
localization. Full curves: potential energy curves of H$_{2}^{+}$.
Dashed lines: initial nuclear wave packet. Vertical lines indicate
transitions between molecular states. The meaning of processes 1,
2 and 3 is explained in the text.}
\end{figure}

The dissociative dynamics leading to eventual electron localization
is well understood \cite{kling2006control}. {A sketch
of this dynamics is shown in Fig. \ref{fig:HHG3D_Process}. During
the first few cycles, the IR field, which is not yet intense enough
to significantly ionize the molecule, induces Rabi-type oscillations between the $1s\sigma_{g}$
and $2p\sigma_{u}$ states, which lie much closer in energy than the
$1s\sigma_{g}$ state and the ionization continuum (process 1). As
a result, excited vibrational states associated
with the $1s\sigma_{g}$ electronic state can be efficiently populated
\cite{PRL_SILVA_2013}, thus creating a vibrational wave packet. A similar vibrational
wave packet is formed in the $2p\sigma_{u}$ state. These 
wave packets then move towards larger internuclear distances (process
2), until they reach a region of internuclear distances where the
$1s\sigma_{g}$ and $2p\sigma_{u}$ electronic states are very close
in energy and are strongly coupled by the IR field. Therefore they mix, leading to localized states $1s\sigma_{g}\pm2p\sigma_{u}$.
The characteristic value of the internuclear distance $R=R_{C}$ where
the onset of localization occurs can be estimated by using the criterion
from \cite{ivanov1993coherent}, $R_{C}E_{0}\omega_{{\rm IR}}\simeq\omega_{gu}^{2}(R_{C})$,
where $E_{0}$ is the field amplitude and $\omega_{gu}$ the energy
difference between the $1s\sigma_{g}$ and $2p\sigma_{u}$ states
at $R=R_{C}$. This criterion yields $R_{C}\simeq3.7$ a.u., shown
in Fig.\ref{fig:HHG3D_Process} with a pink vertical line. Localization
opens the barrier for dissociation, through the process called bond
softening \cite{bucksbaum1990softening}. By then, the IR field has
reached (or nearly reached) its peak intensity and tunnel ionization
becomes prominent (process 3). This enhancement of tunnel ionization
in the long $R$ region is caused by both the reduced ionization potential
(compared to the equilibrium geometry), which decreases with $R$,
and the enhanced ionization caused by electron localization \cite{zuo1995charge,seideman1995role}.
It is important to stress that localization of the electron on one
or the other side of the molecule depends on the phase of the laser
field \cite{kelkensberg2011semi} and, therefore, can be controlled
by the carrier-envelope phase \cite{kling2006control}. Thus, the
ulterior recombination process, i.e., harmonic generation, is acutely
sensitive to the window of $R$ where the medium symmetry is broken.}

\begin{figure}[t]
\begin{centering}
\includegraphics[height=0.8\columnwidth]{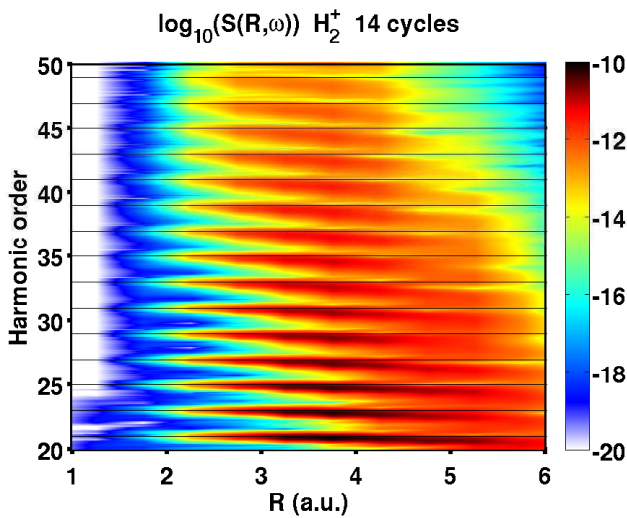} 
\par\end{centering}

\caption{\label{fig:HHG3D_Rdependent}$R$-dependent HHG spectrum obtained
from equation (\ref{R-dep}) for a pulse of $800$ nm, $I=3\times10^{14}$W/cm$^{2}$
and 14 optical cycles for $\hh$. The horizontal lines indicate the
position of odd harmonics.}
\end{figure}


\begin{figure*}[t]
\begin{centering}
\includegraphics[width=0.7\columnwidth]{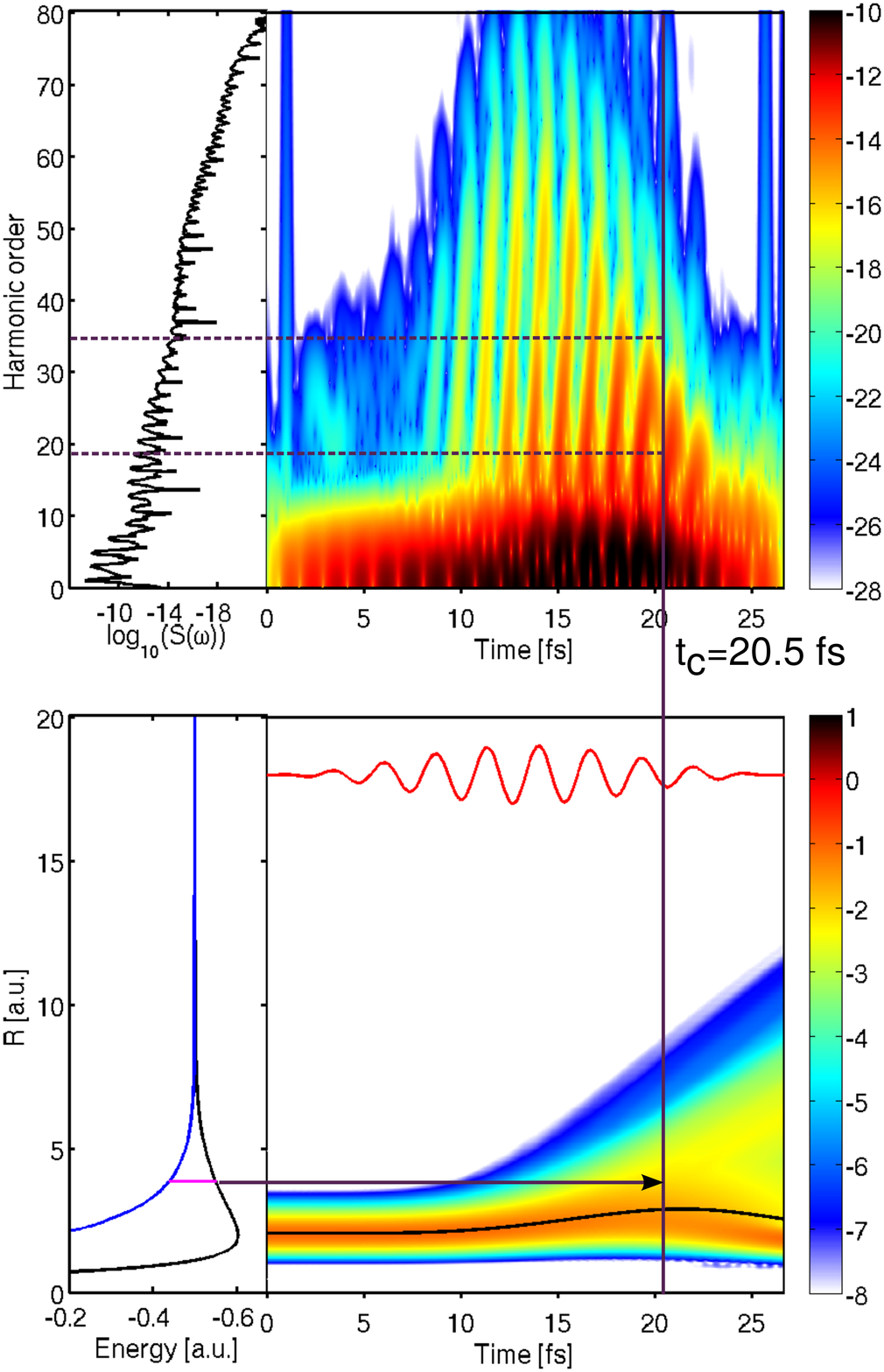}\includegraphics[width=0.7\columnwidth]{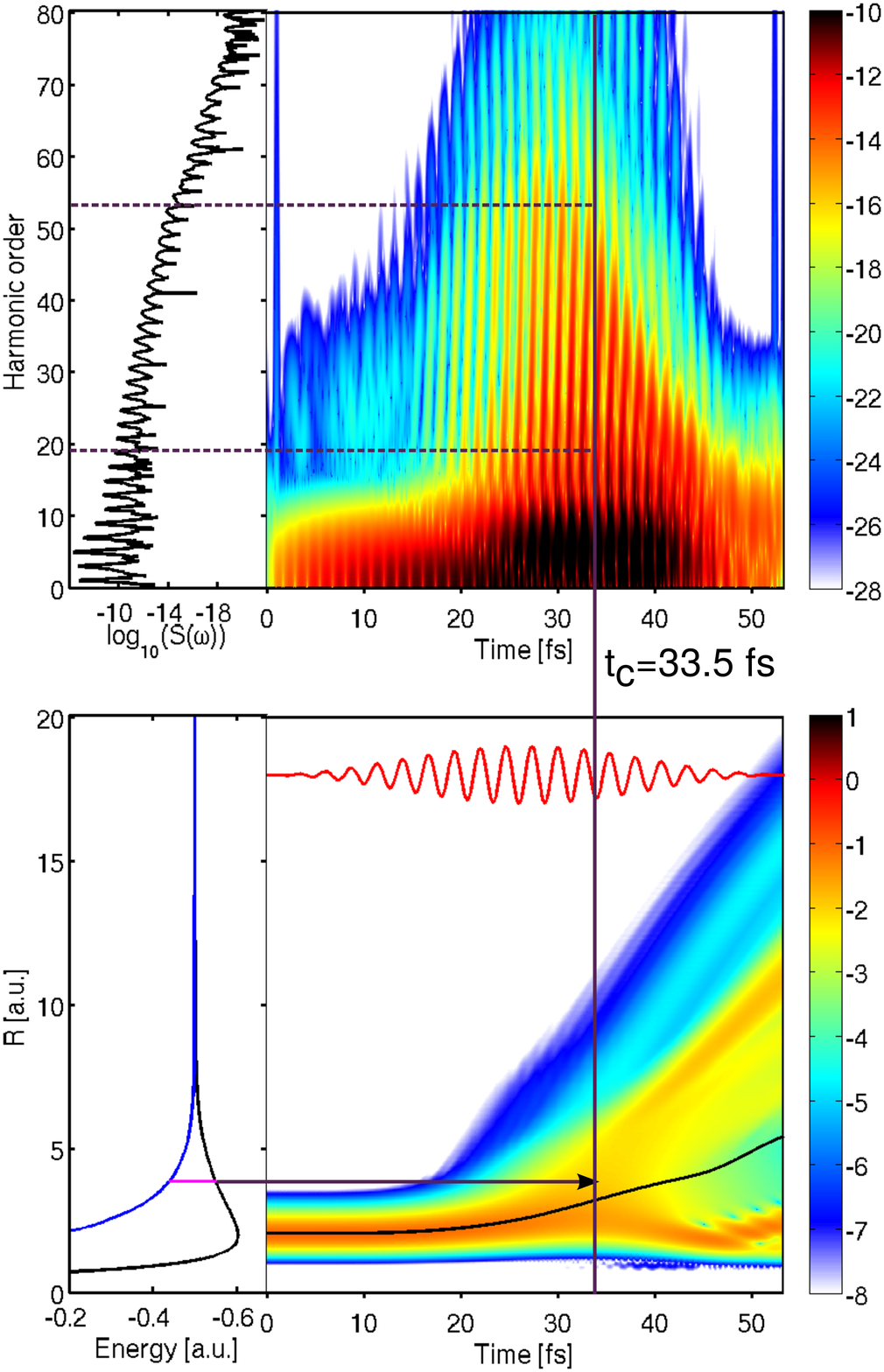} 
\par\end{centering}

\caption{\label{fig:HHG3D_GABOR}Gabor profile (top panels) and nuclear wave
packet distribution (bottom panels) for a pulse of $800$ nm, $I=3\times10^{14}$W/cm$^{2}$
and 10 (left) and 20 (right) optical cycles for $\hh$. The corresponding
HHG spectra are shown on the left of the top panels. The dashed lines
in the top panels represent the ionization threshold (lower line)
and the upper bound for the appearance of even harmonics. For clarity,
the potential energy curves of the H$_{2}^{+}$ states are shown on
the right of the bottom panels. The black horizontal arrow indicates
the value of the internuclear distance at which electron localization
occurs. The vertical lines that goes from the upper to the lower panels
represent the time $t_{C}$ at which electron localization occurs
(see text). The black curves in the lower panel represent the average
$R$ value of the nuclear wave packet.}
\end{figure*}

To quantify this picture and confirm our explanation, we have evaluated
an $R$-dependent HHG spectrum, $\ddot{d}(R,t)$, as the Fourier transform
defined as 
\begin{equation}
\ddot{d}(R,t)=\intop\intop\psi^{*}\left(R,z,\rho,t\right){\cal O}(z)\psi\left(R,z,\rho,d\right)\rho d\rho dz,\label{R-dep}
\end{equation}
{where $\psi$ is the solution of the time-dependent Schr\"odinger equation
(TDSE), $\rho$ and $z$ are cylindrical coordinates describing the position of the electron (see Methods), $R$ is the internuclear distance, and ${\cal O}$ is the dipole acceleration operator.} The calculated
$R$-dependent HHG spectrum is shown in Fig. \ref{fig:HHG3D_Rdependent}
for the 14-cycles pulse. {As can be seen, even harmonics appear near
the localization distance $R\gtrsim R_{C}$ a.u., far beyond the equilibrium distance
$R_{eq}$=1.9 a.u., in agreement with our arguments. This also explains
the lack of even harmonics for heavier isotopes, which dissociate
slower and do not reach the localization region before the pulse is
over.} 

Fig. \ref{fig:HHG3D_GABOR} shows the time-windowed Fourier transforms
(Gabor profiles) and the time evolution of the nuclear wave packets
in $\hh$ for 10 and 20 cycle pulses. For each pulse duration we select
the time, $t_{C}$, at which the density around the critical internuclear
distance, $R_{C}$=3.7 a.u., is largest. By looking at the Gabor profile
at $t>t_{C}$ and at the harmonics that are emitted at $t_{C}$, we
can predict the location of the even harmonics in the HHG spectra.
For the 10-cycles pulse, $R_{C}$ is reached when the intensity of
the laser pulse is already decreasing. This leads to even harmonics
in the lower region of the HHG spectrum. In contrast, for the 20-cycles
pulse, $R_{C}$ is already reached when the intensity of the laser
pulse is still increasing. Consequently, even harmonics appear at
higher energies in the spectrum. {Note that even harmonics arise
for sufficiently long pulses, when the nuclear wave packet has had
enough time to reach $R_{C}$. By controlling the pulse duration one
thus controls the region of the plateau where even harmonics appear.}


\begin{figure}[t]
\begin{centering}
\includegraphics[height=0.7\columnwidth]{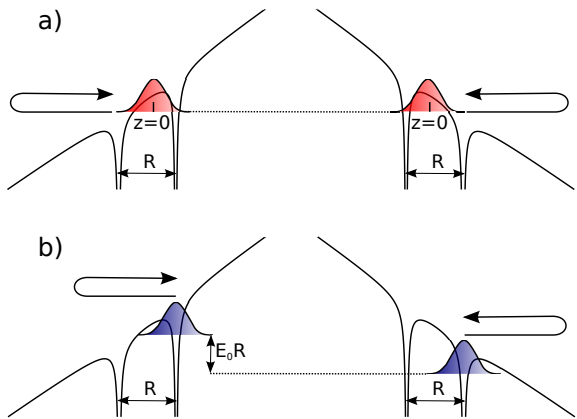} 
\par\end{centering}

\caption{\label{fig:traj}Sketch of the trajectories followed by the electron
in two consecutive half cycles when it is initially delocalized (a)
and localized (b).}
\end{figure}

\begin{figure}[t]
\begin{centering}
\includegraphics[width=1\columnwidth]{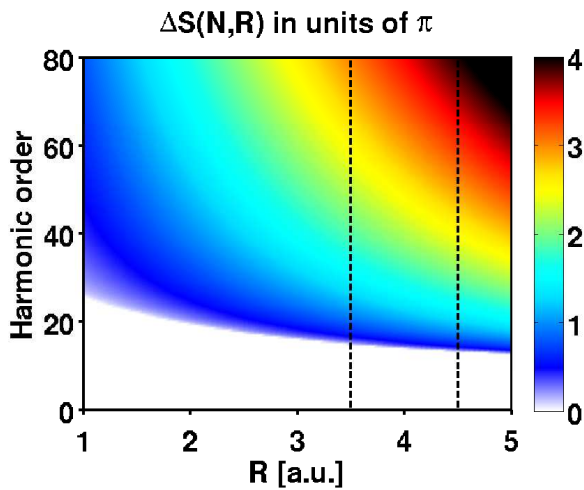} 
\par\end{centering}

\caption{\label{fig:phase}Additional phase difference between two consecutive
half cycles for an electron initially localized on one side of the
molecule (in units of $\pi$). The region between the vertical dashed
lines indicated where localization actually takes place.}
\end{figure}

{This physical picture is further substantiated by analyzing the
phase-shift in the emission bursts during successive half-cycles,
accumulated due to the induced dipole moment. Fig. \ref{fig:traj}
shows a sketch of the typical trajectory followed by an electron starting
in a delocalized and localized initial bound state. 
In the second case, the localized bound state experiences a linear
Stark shift, leading to additional phase difference accumulated between
the ionization $t_{i}$ and recombination $t_{r}$ times, 
\begin{equation}
S_{t_{i}}^{R}=-\int_{t_{i}}^{t_{r}}\frac{R}{2}E\left(t\right)dt.
\end{equation}
For trajectories generated in consecutive half-cycles, the accumulated
phase difference is twice as large, 
\begin{eqnarray}
\Delta S_{t_{i}} & = & S_{t_{i}}^{L}-S_{t_{i}}^{R}=\int_{t_{i}}^{t_{r}}RE\left(t\right)dt=Rv(t_{r})
\end{eqnarray}
In the last equality we have used the three-step model of HHG: for
the electron starting with zero velocity at $t_{i}$, the integral
of the laser field between $t_{i}$ and $t_{r}$ gives its instantaneous
electron velocity $v(t_{r})$ at the return time $t_{r}$. Thus, we
can rewrite this phase difference in terms of the electron recombination
energy $E_{{\rm kin}}=v^{2}(t_{r})/2$, or the emitted photon energy,
$N\omega_{{\rm IR}}$, 
\begin{equation}
\Delta S_{t_{i}}=R\sqrt{2E_{{\rm kin}}}=R\sqrt{2[N\omega_{{\rm IR}}-I_{p}(R)]}.\label{St}
\end{equation}
Fig. \ref{fig:phase} shows this phase difference as a function of
the harmonic order and the internuclear distance. We have checked,
by solving the classical equations of motion numerically, that the
inclusion of the molecular potential adds an extra phase, which is
however much smaller than that shown in Fig. \ref{fig:phase}, so
that the global picture remains unchanged}. In the vicinity of $R_{C}\simeq3.7$
a.u., where localization takes place, there is a large range of harmonic
orders where the additional phase difference between the right and
the left trajectories is $\approx\pi$, thus leading to even harmonic
generation. {According to Fig. \ref{fig:phase}, this
occurs approximately between harmonic orders 20 and 40. Remarkably,
Fig. \ref{fig:phase} predicts a phase difference of $\approx3\pi$,
hence a revival of even harmonic emission, between harmonic orders
60 and 80, in excellent agreement with the appearance of even harmonics in the cut-off region predicted by the ab initio
calculations for H$_{2}^{+}$ (see Fig. \ref{fig:HHG3D_5_cycles}c,d).
Finally, Fig. \ref{fig:phase} also shows that the positions of the
harmonic lines experience a slow frequency shift across the spectrum
as $R_{C}$ is varied, again in agreement with our ab-initio observations.}

\section*{Conclusion\label{sec:Conclusion}}

Using the example of one-electron homonuclear diatomic molecules,
we have shown how dynamics induced in the molecule can lead to a dramatic
breakdown of the standard selection rules in high harmonic generation,
including the nearly complete suppression of odd and the appearance
of even harmonics for multi-cycle laser pulses, in a broad window
of the harmonic spectrum. Our analysis links strong shifts of the
harmonic lines with the appearance of a permanent dipole moment in
dissociating homonuclear molecules, caused by electron localization.
This dipole moment introduces phase shifts between the emission bursts
during successive laser half-cycles, which can approach and exceed
$\pi$. The ultimate origin of symmetry breaking is the sensitivity
of the overall process to the carrier envelope phase of the laser
pulse. 
The fact that minute changes in the driving laser field, associated
with the carrier-envelope phase of a multi-cycle (20-cycle) laser
pulse, can lead to strong effects in the harmonic spectrum, reflects
very strong sensitivity of the underlying dissociation -- localization
dynamics to the details of the driving field. In classical systems,
such extreme sensitivity is characteristic of dynamical chaos. Thus,
our results suggest that high harmonic generation might also be a
sensitive probe for the onset of dynamical chaos in light-driven systems.

\section*{Methods\label{sec:Theoretical-Method}}

Our theoretical method has been described in detail in \cite{Thomas_H2_plus}.
Briefly, we solve the three-dimensional (3D) time-dependent Schr\"odinger
equation (TDSE) in cylindrical coordinates, $\rho$ and $z$ for the
electron, $R$ for the internuclear distance, where $z$ coincides
with the linear polarization direction of the electric field. {We
assume that the molecules are aligned with the linearly polarized
driving IR field, neglecting their rotations. The electron azimuthal
coordinate $\phi$ is removed by the cylindrical symmetry of the problem.
} The TDSE reads 
\begin{equation}
i\frac{\partial\psi{(\rho,z,R,t)}}{\partial t}=[{{H}_{el}(\rho,z,R)+{T}(R)}+{V}(z,t)]\psi{(\rho,z,R,t)},
\end{equation}
where ${H}_{el}={T}_{el}+{V}_{eN}+1/R$ is the electronic Hamiltonian
of H$_{2}^{+}$ or its molecular isotopes, ${T}$ is the nuclear kinetic
energy operator, ${V}={z}{E}(t)$ {describes the interaction with
the laser field} in the length gauge, ${T}_{el}$ is the {nuclear}
kinetic energy operator, {and the Coulomb potential ${V}_{eN}$ describes
the electron -- nuclei interaction. Atomic units are used throughout
unless stated otherwise}. The external laser field is $E\left(t\right)=E_{0}f(t)\sin\left(\omega t\right)$,
where 
\begin{equation}
f\left(t\right)=\begin{cases}
\cos^{2}\left(\frac{\pi t}{T}\right) & \left|t\right|\leq\frac{T}{2}\\
0 & \left|t\right|>\frac{T}{2}\ \ ,
\end{cases}
\end{equation}
{$E_{0}$ the electric field amplitude, $\omega$ the central frequency
corresponding to the wavelength $\lambda=800$ nm, and $T$ the total
pulse duration ranging between 5 and 20 optical cycles ($13.34-53.36$
fs). The peak intensity in all calculations is $3\times10{}^{14}$
W/cm$^{2}$.}


We have {used a non-equidistant cubic 3D grid with} $\left|z\right|<55$,
$\rho<50$ and $R<30$, and grid spacings $\Delta z$=0.1, $\Delta\rho$=0.075
and $\Delta R$=0.05 a.u. at the center of the grid. The grid spacings
increase gradually from the origin to the box boundaries {\cite{Thomas_H2_plus}}.
{The Crank-Nicolson propagator with a split-operator method was used
for time-propagation, with a time step $\Delta t_{elec}$=0.011 a.u.
for the electrons and $\Delta t_{nuc}$=0.11 a.u. for the nuclei.}
The convergence of these parameters was checked. {The initial, ground,
state of the $\hh$ molecule} was obtained by diagonalizing the unperturbed
Hamiltonian {using the SLEPc routines \citep{SLEPc_manual}}. Absorbers
were placed at $\left|z\right|>35$ a.u. and $\rho>30$ a.u. to avoid
{artificial reflections from the boundaries}.

At each time step, we have calculated the time-dependent dipole $\ddot{d}(t)$
as 
\begin{equation}
\ddot{d}(t)=\left<\psi(\rho,z,R,t)|{\cal O}(z)|\psi(\rho,z,R,t)\right>_{\rho,z,R},
\end{equation}
where ${\cal O}$ is the dipole acceleration operator and integration
is performed over electronic ($\rho$,$z$) and nuclear ($R$) coordinates.
The harmonic spectrum $|\ddot{d}(\omega)|^{2}$ is given by the square
of the Fourier transform (FT) of $\ddot{d}(t)$.

\begin{acknowledgments}
We gratefully acknowledge X. B. Bian and A. D.
Bandrauk for fruitful discussions and for sharing their data. This work was accomplished with an allocation of computer
time from Mare Nostrum BSC and CCC-UAM and was partially supported
by the European Research Council Advanced Grant No. XCHEM 290853,
MINECO Project No. FIS2013-42002-R, ERA-Chemistry Project No. PIM2010EEC-00751,
European Grant No. MC-ITN CORINF, European COST Action XLIC CM1204,
and the CAM project NANOFRONTMAG. R.E.F.S. acknowledges FCT - Fundação
para a Ciência e Tecnologia, Portugal, Grant No. SFRH/BD/84053/2012.
M.I. acknowledges support of the EPSRC programme Grant No. EP/I032517/1
and the Voronezh State University. 
\end{acknowledgments}
\appendix

\end{document}